\documentclass[preprint,authoryear,12pt]{elsarticle}

\usepackage{amssymb}
\usepackage{amsmath}
\usepackage{hyperref}
\usepackage{verbatim}

 \biboptions{square, numbers}

\journal{}

\newcommand{\mB}[1]{\mbox{\boldmath{$#1$}}}

\begin{document}

\begin{frontmatter}



\title{Load Forecasting of Supermarket Refrigeration}


\author{Lisa Buth Rasmussen}
\author{Peder Bacher}
\ead{pbac@dtu.dk}
\author{Henrik Madsen}
\address{DTU Compute, Technical University of Denmark, DK-2800 Lyngby, Denmark}

\author{Henrik Aalborg Nielsen}
\address{ENFOR A/S, DK-2970 Hørsholm, Denmark}

\author{Christian Heerup}
\address{Danish Technological Institute, DK-2630 Taastrup, Denmark}

\author{Torben Green}
\address{Danfoss A/S, DK-6430 Nordborg, Denmark}

\begin{abstract}
This paper presents a study of models for forecasting the electrical load for supermarket refrigeration. The data used for building the models consists of load measurements, local climate measurements and weather forecasts. The load measurements are from a supermarket located in a village in Denmark. Every hour the hourly electrical load for refrigeration is forecasted for the following 42 hours. The forecast models are adaptive linear time series models. The model has two regimes; one for opening hours and one for closing hours, this is modelled by a regime switching model and two different methods for predicting the regimes are tested. The dynamic relation between the weather and the load is modelled by simple transfer functions and the non-linearities are described using spline functions. The results are thoroughly evaluated and it is shown that the spline functions are suitable for handling the non-linear relations and that after applying an auto-regressive noise model the one-step ahead residuals do not contain further significant information.
\end{abstract}

\begin{keyword}
refrigeration \sep
load forecasting \sep
numerical weather predictions \sep
adaptive models \sep
base splines \sep
recursive least squares


\end{keyword}

\end{frontmatter}


\section{Introduction}
\label{sec:intro}

Nowadays there is an increased focus on the studies concerning the integration of renewable energy sources into existing energy systems. In 2020 the national goal is that 50\% of the electrical energy consumption in Denmark should be covered by wind energy and hence the development of a Smart Grid is of high priority in Denmark. This development is going to ensure an optimal coherence between the fluctuating energy production from renewables and the energy consumption. The study presented in this paper is carried out as part of iPower, which is a Danish collaboration platform including 32 partners, universities and research institutions as well as industrial companies from various countries. The idea of iPower is to develop an intelligent and flexible energy system, a Smart Grid, that can handle a fluctuating power generation, by enabling increased flexibility in the power load such that it better can follow the generation of wind power and thus decreases the need for grid and reserve capacity investments. The ability for the power consumers to contribute with flexibility is a key issue and one of the challenges is to create application-configurable control schemes for the industrial consumption side. An industrial consumer, that could provide flexibility is the Danish supermarkets, they are using a large amount of electricity on lightning, cashiers, refrigerators and coolers, etc. In many of the proposed Smart Grid setups in the literature load forecasting is an important part of the control schemes \citep{Fang2012}. \\

Many approaches to load forecasting are found in literature, however no studies were found which specifically consider forecasting of load for refrigeration. For load forecasting different kinds of data-driven modelling methods are used from parametric models, such as ARIMA and ARMAX models, and over to fully non-parametric methods, such as Support Vector Machines (SVM) and Artificial Neural Networks (ANN). Regression based methods are used by \citet{Chen1995}, and \citet{Charytoniuk1998} and \citet{Fan2012} propose a semi and non-parametric regression models. \citet{Penya2011} compare an ARIMA model with ANNs for forecasting of electricity load for a HVAC system for an office building. \citet{Fard2013} presents a combined method based on several of the commonly used methods. A good overview of references with ANNs approaches to load forecasting is given by \citet{Hippert2001}, and \citet{Datta} apply an ANN model for prediction of the total electrical load for a supermarket. \citet{Halvgaard2012} use heat load forecasting as part of an economic Model Predictive Control (MPC) for providing demand response using a hot water tank for thermal energy storage, in the same way load forecasting for refrigeration can be used with an MPC and an ice storage tank, see \citep{Henze2004}. The subject of flexible power consumption in refrigeration systems has a huge potential, see for example \cite{Gybel} and \cite{Shafiei2013}.\\

This paper presents a study of models for forecasting the electrical load from supermarket refrigeration. The data used for building the forecast models are hourly load measurements, local measured ambient temperature and numerical weather predictions (NWPs) for a summer period of 3 months (May, June, July). The forecast models are adaptive linear time series models which are fitted with a computationally efficient recursive least squares scheme (RLS). Every hour the hourly load for refrigeration for the following 42 hours is forecasted. The dynamic relations between the inputs and the load are modelled using linear transfer functions and non-linearities are handled with spline functions. The refrigeration system operates in two regimes; one during opening hours and another during closing hours, which is modelled by a regime switching model. Different approaches to predicting the regime is tried; one simply by using the fixed opening and closing hours and another more automatized approach where a forecasted diurnal curve is used. The results are thoroughly evaluated and discussed. Finally, ideas for further modifications are suggested and conclusions are drawn.

\section{Data}
\label{sec:data}
The data used in the study consists of measurements from a supermarket located in a village in Denmark. The local measured load and ambient temperature are used together with NWPs.

\subsection{Electrical load for refrigeration and local temperature measurements}
\label{subsec:dat:observations}
\begin{figure*} 
   \includegraphics[scale=1]{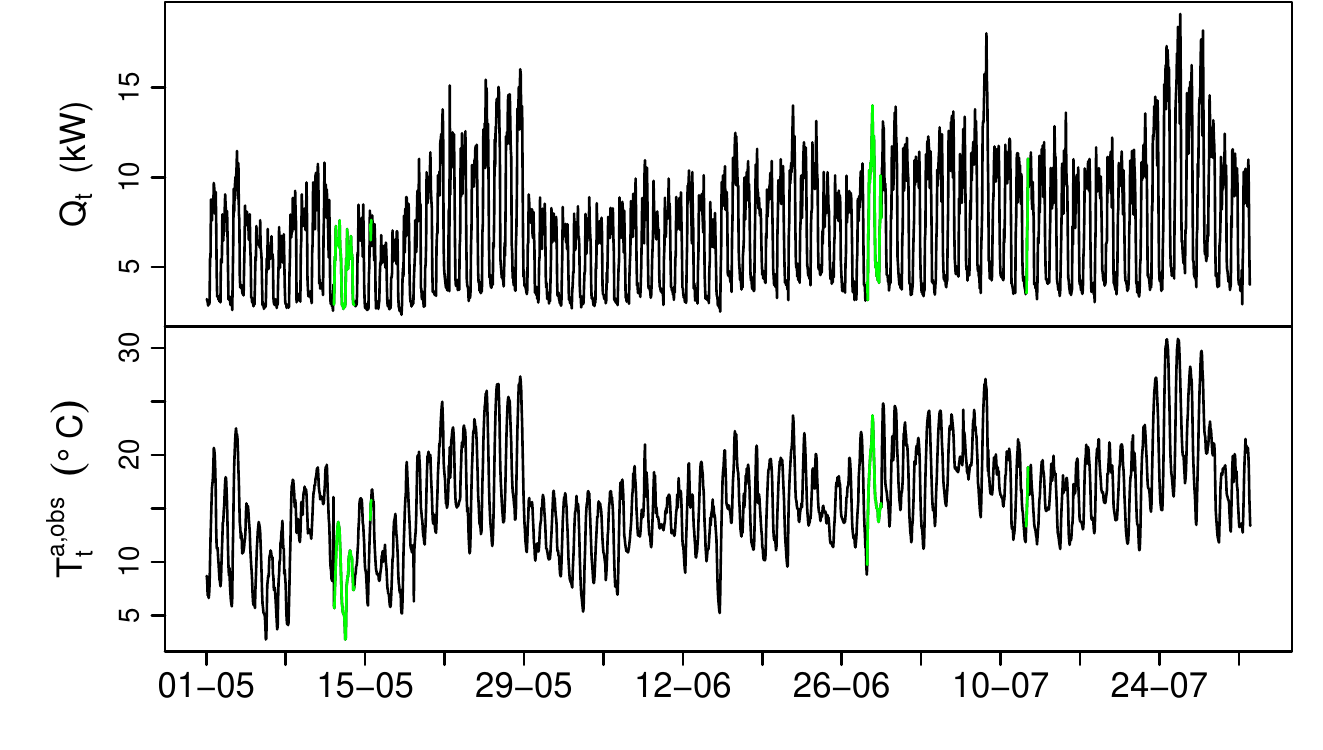} 
   \caption{Time series plot with hourly average values of the load and the local ambient temperature. The green parts are the replaced values}
   \label{fig:data}
\end{figure*}

\begin{figure} 
   \includegraphics[scale=1]{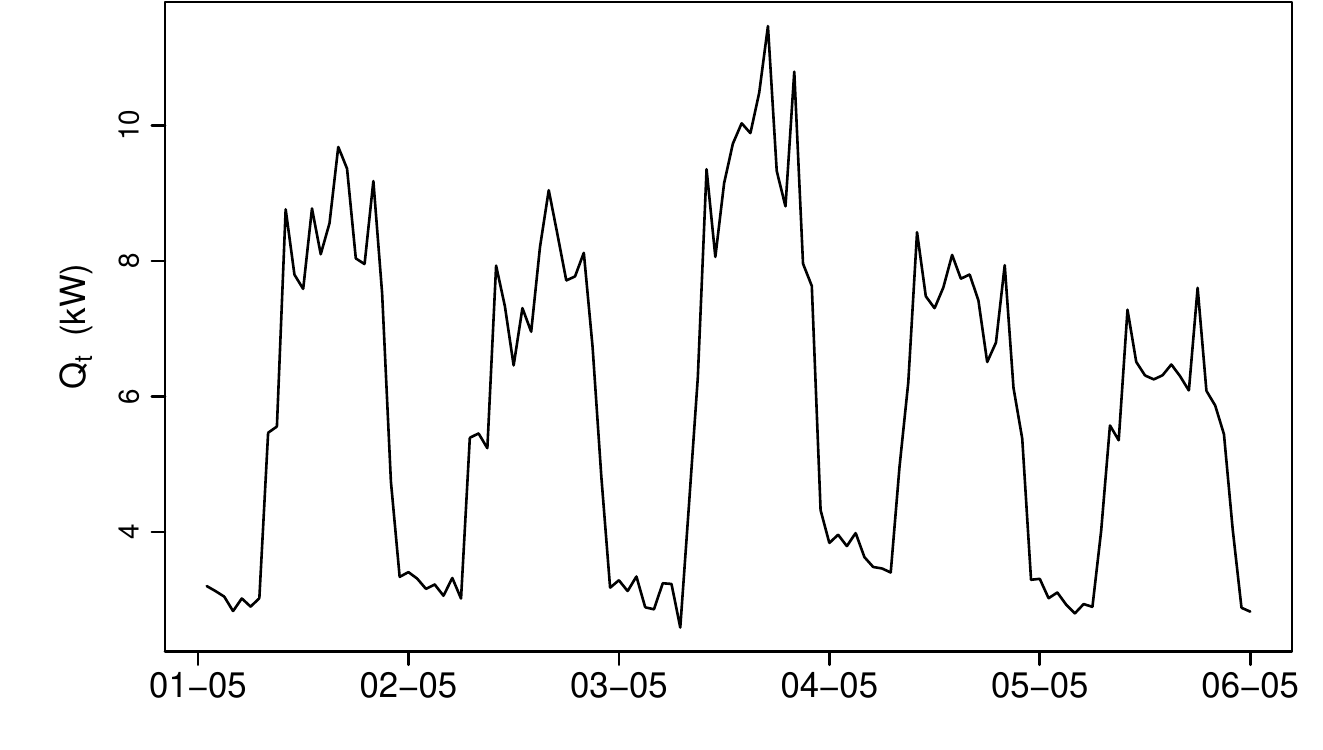} 
   \caption{Time series plot of hourly average values of the load for five days}
   \label{fig:data1}
\end{figure}

The load measurements are the electrical load of the compressors of a trans critical CO2 refrigeration system which provides cooling to four low and seven medium temperature cooling units. The period acquired is from May 1$^{st}$ to August 1$^{st}$ 2012. The measured time series are hourly averaged values in $kW$ and are denoted by
$$
\{ Q_t ,\quad t=1,\ldots,N\}
$$
Local measured ambient temperature is also hourly average values, in $^\circ C$, denoted by
$$
\left\lbrace T^{a,obs}_{t}, \quad t=1,\ldots,N\right\rbrace
$$
where $N=2208$. A few small parts of the time series are missing. Two gaps, first approx. two days long, and second approx. one day long, are replaced with data, corresponding to the same hours and weekdays, from the week before. Smaller gaps are replaced with the corresponding hour from the day before. In the gaps other corresponding series (ambient temperature and NWPs) are also replaced. These simple replacements are limited and will therefore only affect the results marginally. A plot of data is found in Figure \ref{fig:data}, where the replaced gaps are displayed in green. In Figure \ref{fig:data1} a period of five days is plotted. It is clearly seen that the system operates in two regimes, which are identified as closing and opening hours, where at closing hours the load is lower than opening hours. This is mainly because the supermarket is closed at night, and the cabinets without doors are covered by isolation material. Also high frequency peaks are seen in the opening hours, which could be related to defrosting of the low temperature cabinets, that is scheduled in the morning and evening, individually for the different types of low temperature cabinets. For more detailed information about the system, see \citep{ESO}.

\begin{figure} 
   \includegraphics[scale=1]{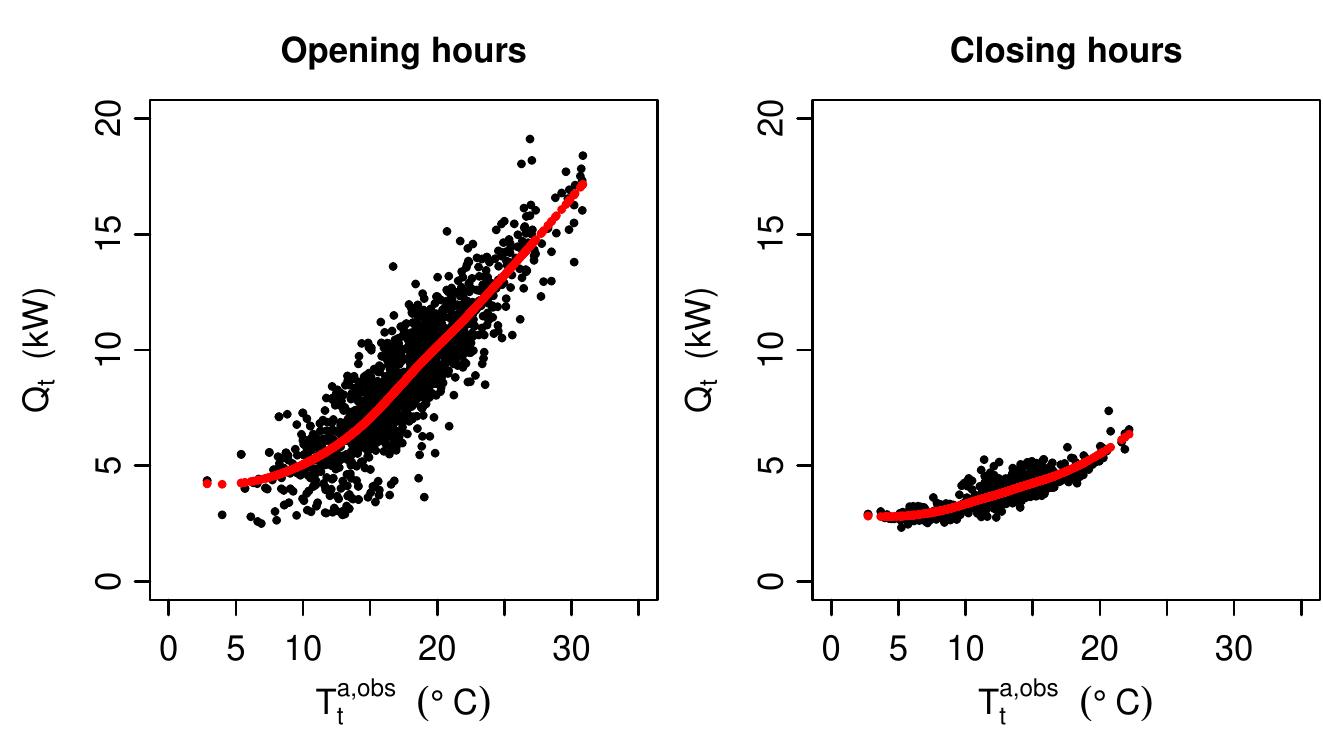} 
   \caption{Ambient temperature versus load, for opening hours (left) and closing hours (right). The red lines are smoothed local second order regression estimates fitted using the R \cite{RCoreTeam2013} function loess with default values.}
   \label{fig:data2}
\end{figure}

In Figure \ref{fig:data2}, the load and the local ambient temperature are plotted against each other, together with a smooth local regression estimate \cite{Cleveland1992}. The opening hours for the supermarket are all day 8 am to 9 pm and the period is separated from 7 am to 10 pm as the opening hours, which was found as the best interval for the separation by visual inspection of plots. The plots shows the positive dependency between the load and the local ambient temperature. The main cause of the increasing load when the ambient temperature raise, is that the compressors need to work more to increase the pressure of the refrigerant gas, in order to transfer the needed amounts of thermal energy to the surroundings. The smoothed local estimates (red lines) indicates of a non-linear dependency, which should be taking into account by the forecast models. 

\subsection{Numerical weather predictions}
The numerical weather predictions (NWPs) used in the models are provided by Danish Meteorological Institute (DMI). The model used by DMI, to make the NWPs, is DMI-HIRLAM-S05, which has a 5 kilometers grid and 40 vertical layers \citep{DMI}. The NWPs consist of climate variables, with hourly values of the forecasts, which are updated four times per day and takes 4 hours to complete, i.e. the forecast starting at 06:00 is available at 10:00.\\

The predicted ambient temperature has the unit $^\circ C$ and the $k^{th}$ horizon is denoted by
\begin{align}
\left\lbrace T_{t+k|t}^{a,nwp}, \quad t=1,\ldots,N,\quad k=1,\ldots,42\right\rbrace
\end{align}
where $N=2208$.

\subsection{Combining local observations with NWPs}
When a load forecast is calculated at time $t$, past values of the model inputs are being used, therefore the local measured ambient temperatures are combined with the NWPs. The combining is achieved by forming a time series at time $t$, for a specific horizon $k$, denoted
\begin{align}
\left\lbrace T_{t+k|t}^{a}\right\rbrace = \left\lbrace \ldots,T_{t-1}^{a,obs}, T_{t}^{a,obs},T_{t+1|t}^{a,nwp},T_{t+2|t}^{a,nwp},\ldots,T_{t+k|t}^{a.nwp}\right\rbrace
\end{align}
The combined series are used to include dynamics in the model in an efficient way by low-pass filtering as explained in Section \ref{subsec:mod:Lowpass}.\\

The NWP predicted ambient temperature $T_{t+1|k}^{a,nwp}$ represents the temperature for the local area where the supermarket is located and compared to the measured ambient temperature a bias is observed, approximately between 1 to 5 $^\circ$C below the measured temperature. The forecasts are therefore calibrated to the local measurements with a simple local polynomial regression model.

\section{Methods}
\label{sec:models}
The applied models are formed using the same techniques as \citep{Nielsen2000} for forecasting of the total heat load for many houses and by \citet{Bacher2012} for forecasting of the heat load for single family houses. The models are based on prior physical knowledge of the heat dynamics, which is coupled with statistical time series modelling techniques. This allows for inclusion of heat transfer effects related to the climate variables in combination with a time adaptive estimation scheme applied to meet changing conditions or variables not directly accounted for by the model, for example the effect of the delivery of new goods placed in cabinets. In order to describe the systematic patterns in the load, a diurnal curve and regime switching models are used, furthermore the non-linear correlation between load and temperature is described using splines. Two parameters of the forecasting models are estimated by minimizing the root mean square error (RMSE) in an off-line setting, they are related to slow varying features of models, as explained below. The forecast models are fitted separately for each forecast horizon $k$, which means that the same model formulation - i.e. same inputs and model structure - is used, only the parameters and coefficient values are different for each horizon. By this approach the increasing uncertainty related to the forecast horizon is taken into account.
\subsection{Time adaptive models}
\label{subsec:mod:Timeadap}

The models are fitted with a $k$-step recursive and adaptive least squares scheme \citet{Madsen2008}. This means that the coefficients in the model can change over time and adapt optimally to changing conditions in a least squares sense, hence if the residuals are Gaussian white noise (identical and independently normal distributed) then the predictions are based on the maximum likelihood estimates of the parameters. The coefficients are recursively updated, which means that only a few matrix operations are required to compute an updated forecast, hence the scheme is computationally very fast.\\

The RLS scheme used to update the coefficients is
\begin{align}\label{eq:rls}
\mathbf{R}_t &= \lambda\mathbf{R}_{t-1} + \mathbf{X}_{t-k}\mathbf{X}_{t-k}^T\\
\mB{\hat{\theta}}_t &= \mB{\hat{\theta}}_{t-1} + \mathbf{R}_t^{-1}\mathbf{X}_{t-k} [Y_t-\mathbf{X}_{t-k}^T\mB{\hat{\theta}}_{t-1}]\nonumber
\end{align}
It is a recursive implementation of a weighted least squares estimation, where the weights are exponentially decaying over time. With $\mathbf{X}_t$ as the regressor vector, $\mB{\theta}_t$ as the coefficient vector and $Y_t$ is the dependent variable (observation at time $t$), the $k$-step prediction at $t$ is 
\begin{align}\label{eq:rlspredict}
\widehat{Y}_{t+k|t} = \mathbf{X}_t^T\mB{\hat{\theta}}_t
\end{align}
A single parameter is required, the forgetting factor $\lambda$, for describing how fast historical data is down-weighted. The weights are equal to
$$
w(\Delta t) = \lambda^{\Delta t}
$$
where $\Delta t$ is the age of the data in hours. This implies that for $\lambda = 0.95$ the weights are halved in 13 hours and for $\lambda = 0.998$ they are halved in 346 hours ($\sim$ 14 days). Further details about the algorithm are found in \citep{SolPeder}.

\subsection{Diurnal curve}
\label{subsec:mod:Diur}
A diurnal curve is included in the models for describing systematic diurnal patterns in the load. The curve is modelled as a harmonic function using the Fourier series
\begin{align}\label{eq:diurnal}
\mu(t,n_{har},\mB{\alpha}_{diu}) &= \sum_{i=1}^{n_{har}} \alpha^{wor}_{1,i} \sin\left(\frac{t_{tod} i \pi }{ 12 } \right) I_t + \alpha_{2,i}^{wor}\cos\left(\frac{t_{tod} i \pi }{ 12 }\right) I_t \\
&+ \sum_{i=1}^{n_{har}} \alpha^{wee}_{1,i} \sin\left(\frac{t_{tod} i \pi }{ 12 }\right) \left(1-I_t\right) + \alpha_{2,i}^{wee}\cos\left(\frac{t_{tod} i \pi }{ 12 }\right) \left(1-I_t\right)\nonumber
\end{align}
where $t_{tod}$ is the time of day in hours at time $t$, $n_{har}$ is the number of harmonics included in the Fourier series, 
\begin{align}
\mB{\alpha}_{diu}=(\alpha^{wor}_{1,1},\dots,\alpha^{wor}_{1,n_{har}},\alpha^{wor}_{2,1},\dots,\alpha^{wor}_{2,n_{har}},\alpha^{wee}_{1,1},\dots,\alpha^{wee}_{1,n_{har}},\alpha^{wee}_{2,1},\dots,\alpha^{wee}_{2,n_{har}})  
\end{align}
is a vector consisting of the coefficients for the included harmonics and
\begin{align}
  I_t = \left\{\begin{array}{rl}
  1 & \text{for workdays} \\
  0 & \text{for weekend} \\
\end{array}\right.
\end{align}
is an indicator time series switching between a diurnal curve for workdays and another for weekends. Note that $\mB{\alpha}_{dia}$ are the coefficients which are fitted with the RLS scheme, i.e. they become part of $\mB{\theta}_t$ in Equation \eqref{eq:rls}, and they could be denoted with a $t$ since they are time varying, however this is left of for simplicity of writing. In the present study 10 harmonics are included in order to be able to match the almost square shaped daily load pattern.

\subsection{First order low-pass filtering}
\label{subsec:mod:Lowpass}
The main effect causing the positive dependence between the load and ambient temperature is the increased work needed from the compressors to increase the pressure of the refrigerant gas. This should lead to a fast response (on an hourly time scale) from the ambient temperature to the load. However the refrigeration system consists in a addition to the compressor rack of piping and cabinets, and it is interacting thermally with the building and its surroundings, therefore some low-pass filtering could also be expected. The heat dynamics of a passive thermal system, e.g. a building, can be described by lumped parameter RC-models, see for example \cite{Braun2002} and \cite{Jimenez2008}, which correspond to rational transfer functions.  The response in the load to changes in the climate variables is modelled with rational transfer functions. The simplest first order low-pass filter, with a stationary gain of one, is a model of the system heat dynamics formed by an RC-model with a single resistor and a single capacitor. As an example the transfer function from the ambient temperature to the load is modelled with
\begin{align}
\label{eq:lowpass}
Q_t = \alpha_aH_a(q)T_t^a
\end{align}
where
\begin{align}
H_a(q) = \frac{1-a_{T_a}}{1-a_{T_a}q^{-1}}
\end{align}
and where $q^{-1}$ is the backward shift operator ($q^{-1}x_t=x_{t-1}$), $\alpha_a$ is the stationary gain from the ambient temperature to load and $a_{T_a} \in [0,1]$ is a parameter which corresponds to the time constant for the part of the system affected by changes in ambient temperature. If the system has a high thermal mass and good insulation, a relatively high $a_{T_a}$ is expected, thus the filter parameter needs to be adapted to the particular system in order to describe the dynamics properly.

\subsection{Splines}
\label{subsec:mod:bspline}
The non-linear relation between ambient temperature and load, described in Section \ref{subsec:dat:observations} is modelled using splines. A spline function is a piecewise-defined smooth polynomial function of order $p$ with a sequence of knot points $\mathbf{z} = (z_1,z_2,\dots,z_n)$ and it has a continuous derivative up to order $p-1$ at the knot points \citep{deBoor}. A spline can be formed by a linear combination of basis splines (B-splines)
\begin{align}
S_{p,\mathbf{z}}(x) = \left\lbrace \sum_{j=1}^n \beta_j B_{j,p,\mathbf{z}}(x)\quad ,\beta_j \in \mathbb{R} \right\rbrace.
\end{align}
where $x$ is the input. To fit a spline function for two data series a linear regression model can be applied
\begin{align}
Y_i = \sum_{j=1}^n \beta_j B_{j,p,\mathbf{z}}(X_i) + \epsilon^{spl}_{i}
\end{align}
where the coefficients $\beta_j$ and B-splines then forms a spline function. The knots are where the polynomial pieces connect. The amounts and placement of the knots needs to be considered: too few knots makes the spline model biased and too many knots makes the splines too varying with the possibility of over-fitting to data.

\section{Models}
In this study three models are presented. The first model is the best linear model identified in \citet{ESO2} and the two other models are modifications of the first model for better prediction of the regimes and for improving the description of the non-linear effect of the ambient temperature. In this section the models are presented and in the next section the results are presented and analysed. The models are in general presented as
\begin{align}
\label{eq:modelbasic}
Q_{t+k} = \widehat{Q}_{t+k|t} + \epsilon_{t+k}
\end{align}
where $Q_{t+k}$ is the load at time $t+k$, $\widehat{Q}_{t+k|t}$ is the forecasted value available at time $t$ and $\epsilon_{t+k}$ is the residual for forecast horizon $k$.

\subsection{Root Mean Square evaluation}
\label{subsec:mod:RMSE}
To evaluate the models the root mean square error (RMSE) for the $k^{th}$ horizon is used. The RMSE$_k$ is defined as
\begin{align}
RMSE_k = \sqrt{\frac{1}{N}\sum_{t=1}^{N}\epsilon^{2}_{t+k}}
\end{align}
where $k=1, \ldots. 42$. The period before May 15$^{th}$ will be used as a burn-in period and is therefore excluded from the RMSE$_k$ calculations.

\subsection{Parameter optimization}
\label{subsec:mod:paramopt}
As described above several parameters need to be optimized for each horizon. The optimization is carried out in an off-line setting by minimizing the RMSE$_k$ for each horizon $k = 1,\ldots,42$ separately. The function \verb|optim()| from the R-software \citep{RCoreTeam2013} is used for the optimization. \\

The following parameters are optimized:
\begin{itemize}
\item The forgetting factor: $\lambda$
\item The coefficient for the low-pass filtering of ambient temperature: $a_{T_a}$
\end{itemize}
These parameters describe properties of the system, which changes very slowly in time, and hence their optimization do not have the need to be updated very often, e.g. once per month is most likely sufficient. The properties of the optimization is not studied in further details in this work.

\subsection{Fixed regime and linear model}
The first model is denoted by $Model_{\mathit{fix.lin}}$. The model includes a diurnal curve, which switches for workdays and weekends as described in Section \ref{subsec:mod:Diur}, and the effect of ambient temperature is included by letting the ambient temperature enter through a low-pass filter and then switched between the opening and closing hours regimes at two fixed times of the day. The forecasted value is
\begin{align}\label{eq:fixedlinear}
\widehat{Q}_{t+k|t} &= \mu(t+k,n_{har},\mB{\alpha}_{diu}) + \alpha_{iop} \; I_{t+k|t}^{fix}  + \alpha_{aop} \;  I_{t+k|t}^{fix} H_a(q)T_{t+k|t}^{a} \\
&+ \alpha_{icl} \left(1-I_{t+k|t}^{fix}\right) + \alpha_{acl} \left(1-I_{t+k|t}^{fix}\right)H_a(q)T_{t+k|t}^{a} 
\end{align}
where $\mu(t+k,n_{har},\mB{\alpha}_{diu})$ is the diurnal curve, the indicator time series
\begin{align}
  I_{t+k|t}^{fix} = \left\{\begin{array}{rl}
  1 & \text{for } 8 \leq (t+k)_{tod} \leq 21 \\
  0 & \text{for } (t+k)_{tod} < 8 \lor 22 < (t+k)_{tod} \\
\end{array}\right.
\end{align}
where $(t+k)_{tod}$ is the time of day in hours and time $t+k$, switches between the opening and closing regime, $H_a(q)T_{t+k|t}^{a}$ is the low-pass filtered ambient temperature. The coefficients $\alpha_{iop}$ and $\alpha_{aop}$ are the intercept and slope for opening hours, and similarly $\alpha_{icl}$ and $\alpha_{acl}$ are for the closing hours. When the model is fitted with the RLS scheme it is these four coefficients, which, together with $\mB{\alpha}_{diu}$, forms the coefficient vector $\mB{\theta}_t$.

\subsection{Variable regime and linear model}\label{sec:vari-regime-line}
The second model introduced is denoted with $Model_{\mathit{var.lin}}$. The forecasted value is
\begin{align}\label{eq:variablelinear}
\widehat{Q}_{t+k|t} &= \mu(t+k,n_{har},\mB{\alpha}_{diu}) + \alpha_{iop} \; I_{t+k|t}^{var} + \alpha_{aop} \; I_{t+k|t}^{var}H_a(q)T_{t+k|t}^{a}  \\
&+ \alpha_{icl} \left(1-I_{t+k|t}^{var}\right) + \alpha_{acl} \left(1-I_{t+k|t}^{var}\right) H_a(q)T_{t+k|t}^{a} 
\end{align}
where only the prediction of opening and closing regime is different. To make the regime switching more adaptive, then, instead of relying on a fixed time interval for the opening and closing regimes, which could be manually provided, a predicted diurnal curve is used. The model applied for predicting the regime is
\begin{align}
  Q_{t+k} = \mu(t+k,n_{har},\mB{\alpha}_{t+k}^{rgm}) + \alpha_{int} + \alpha_{a} H_a(q)T_{t+k|t}^{a} + \epsilon_{t+k}^{rgm}
\end{align}
The model is fitted and the estimated coefficients are used to predict the regime by
\begin{align}
  I_{t+k|t}^{var} = \left\{\begin{array}{rl}
  1 & \text{for } \mu(t+k,n_{har},\mB{\widehat{\alpha}}_{t}^{rgm}) \geq 0 \\
  0 & \text{for } \mu(t+k,n_{har},\mB{\widehat{\alpha}}_{t}^{rgm}) < 0 \nonumber
\end{array}\right.
\end{align}
Note that the coefficients $\mB{\widehat{\alpha}}_{t}^{rgm}$ are estimated separately for each horizon $k$ using the RLS scheme in Equation \eqref{eq:rls} and that at the time $t$ of prediction the currently available estimates are used as in Equation \eqref{eq:rlspredict}.

\subsection{Variable regime and non-linear model}
The third model is denoted by $Model_{\mathit{var.nonlin}}$. It is extended further from the second model by modelling the non-linear effect for the ambient temperature, which was found by considering the plot in Figure \ref{fig:data2}. The forecasted value is
\begin{align}
\label{eq:modelspoff}
\widehat{Q}_{t+k|t} &= \mu(t+k,n_{har},\mB{\alpha}_{diu}) + \alpha_{iop} \; I_{t+k|t}^{var}  + \alpha_{aop} \; I_{t+k|t}^{var}S_{t+k|k}^{open} \\
&+ \alpha_{icl}  \left(1-I_{t+k|t}^{var}\right) + \alpha_{acl}  \left(1-I_{t+k|t}^{var}\right) S_{t+k|k}^{close}
\end{align} 
where the ambient temperature is first low-pass filtered and then, before included in the model as input, modelled with a spline function for each regime
\begin{align}
\label{eq:splinefct}
S_{t+k|k}^{open} =& \sum_{j=1}^n \hat{\beta}_{j}^{open} B_{j,p}^{open}\big(H_a(q)T^{a}_{t+k|t}\big) & \text{for } \mu(t+k,n_{har},\mB{\hat{\alpha}}_{t}^{rgm}) \geq 0\\
S_{t+k|k}^{close} =& \sum_{j=1}^n \hat{\beta}_{j}^{close} B_{j,p}^{close}\big(H_a(q)T^{a}_{t+k|t}\big) & \text{for } \mu(t+k,n_{har},\mB{\hat{\alpha}}_{t}^{rgm}) < 0
\end{align}
where the opening and closing regimes are predicted the same way as described above in Section \ref{sec:vari-regime-line}. An order of $p=3$  and number of knots $n=5$ are found reasonable based on visual inspection of plots and the knots sequence are simply taken as the equally distributed quantiles of the ambient temperature. This kind of two-stage approach, where non-linear effects are modelled in an off-line setting before a linear on-line model, is widely used, for example in wind power forecasting \citep{Wind}. As described in Section \ref{subsec:mod:bspline} a linear regression model is applied to find the estimated parameters $\hat{\beta}_{j}^{open}$ and $\hat{\beta}_{j}^{close}$ for the spline functions, here using on the observed load for the whole period for each regime separately. This is not operationally possible, due to use of the future observations, however due to the lack of data for several years and since only a few degrees of freedom are used (the number of parameters for the splines), the performance will only be marginally lower if historical data from previous years were used instead.

\section{Results}
\label{sec:results}

In this section the results from forecasting with the described models are presented and evaluated. First the forecast performance for all three models are evaluated and then residuals for the best performing model is further analyzed.\\

\begin{figure} 
   \includegraphics[scale=1]{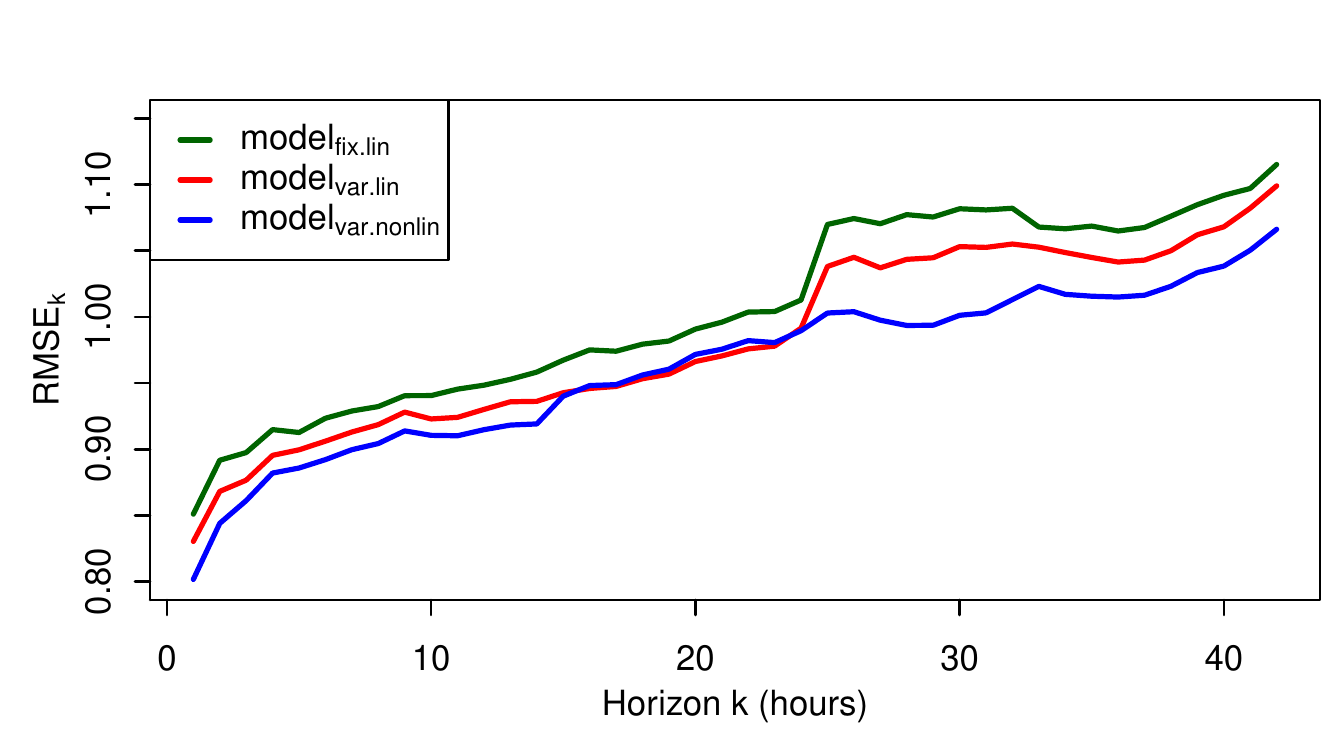} 
   \caption{The RMSE$_k$ for the models for each horizon $k$}
   \label{fig:rmse}
\end{figure}
The RMSE$_k$ of the residuals is calculated for each horizon $k$ and plotted for each model in Figure \ref{fig:rmse}. It is seen that using the adaptive regime prediction in $Model_{\mathit{var.lin}}$ the forecasting performance clearly increases compared to the fixed time of day regime used in $Model_{\mathit{fix.lin}}$. It is also seen that including the non-linear effects of ambient temperature improves the performance, especially for the longer horizons (above 25-steps ahead). The linear models can include non-linear effects by adapting the diurnal curve on the horizons shorter than 24 hours, but not on longer horizons. Hence the $Model_{\mathit{var.nonlin}}$ is found to be the most suitable model and the residuals from this model are analysed in the following. 

\begin{figure} 
   \includegraphics[scale=1]{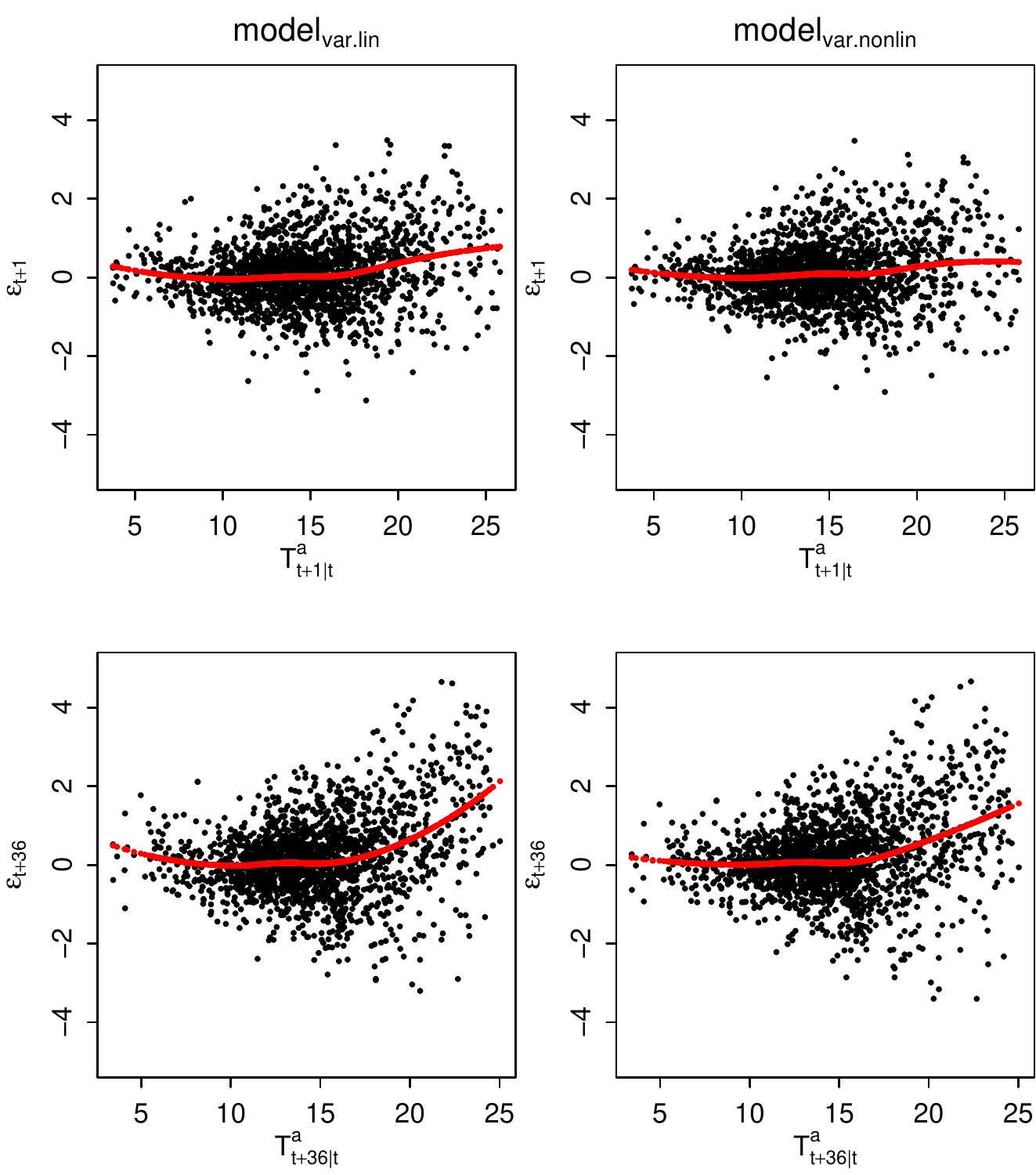} 
   \caption{Plot of residuals versus ambient temperature for 1-step horizon (top) and 36-steps horizon (bottom), to the left $Model_{\mathit{var.lin}}$, to the right $Model_{\mathit{var.nonlin}}$}
   \label{fig:resta}
\end{figure}
In Figure \ref{fig:resta} the 1-step and 36-steps ahead residuals for $Model_{\mathit{var.lin}}$ and $Model_{\mathit{var.nonlin}}$ are plotted versus the ambient temperature input including smoothed local regression estimates. The plots reveals that for $Model_{\mathit{var.lin}}$ the residuals are biased for levels of the ambient temperature above approximately 17-20 $^\circ$C, as also found by \citet{ESO2}. Comparing with $Model_{\mathit{var.nonlin}}$ and taking into account that the linear models are time adaptive and therefore do adapt to the non-linear effects over time, it is seen that the description of the non-linear effect from ambient temperature decreases the dependencies, however not entirely.\\

\begin{figure} 
   \includegraphics[scale=1]{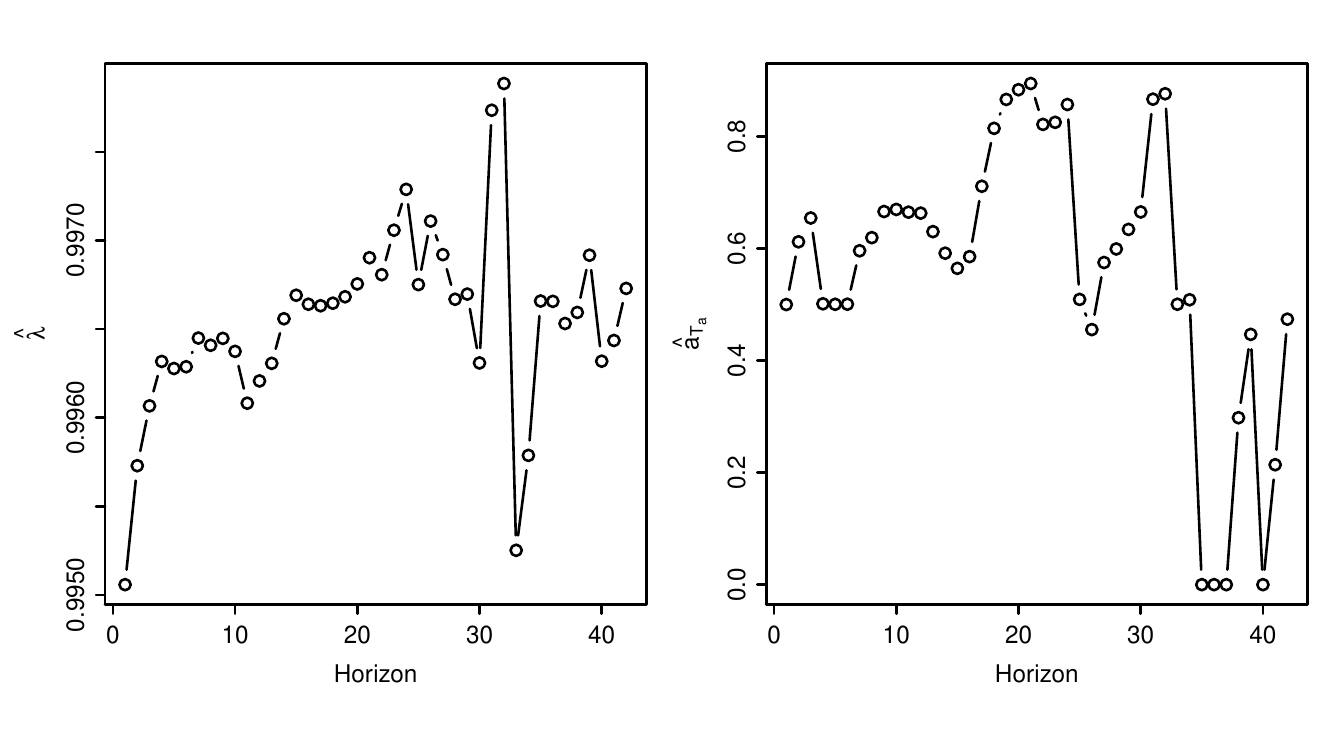}
\vspace{-0.5cm}
   \caption{The two parameters which are optimized off-line for each horizon. The left plot is of the forgetting coefficient and the right plot is of the low-pass filtering coefficient.}
   \label{fig:prm}
\end{figure}
The optimized values of the two parameters optimized off-line, see Section \ref{sec:models},  are plotted versus the horizon in Figure \ref{fig:prm}. The optimized forgetting factor $\hat{\lambda}$ changes increases for the first few horizons from around 0.995 (weighting of data is halved in approx. 6 days) up to around 0.9965 (weighting halved in approx. 8 days) where it more or less stays for the longer horizons, with some variations around the 32 hours horizon. These values are found reasonable and shows that the models does adapt quite fast to changing conditions. The optimized coefficient $\hat{a}_{T_a}$ of the low-pass filter from the ambient temperature has up to the 16 hours horizon a value around 0.6, which indicates a fast response to the ambient temperature. For longer horizons it goes up to around 0.85 and finally drops to 0 at the 35 hours horizon. This variation is quite high and indicates, together with the fast response, that the low-pass filtering doesn't have a huge influence on the forecasts for the refrigeration system. This is also underpinned when compared to the similar forecast models for heat load in buildings, where the coefficient was found to be around 0.95 \cite{Bacher2012} with much less variation, i.e. a much slower response and hence more low-pass filtering effect from the ambient temperature for buildings.

\begin{figure} 
   \includegraphics[scale=1]{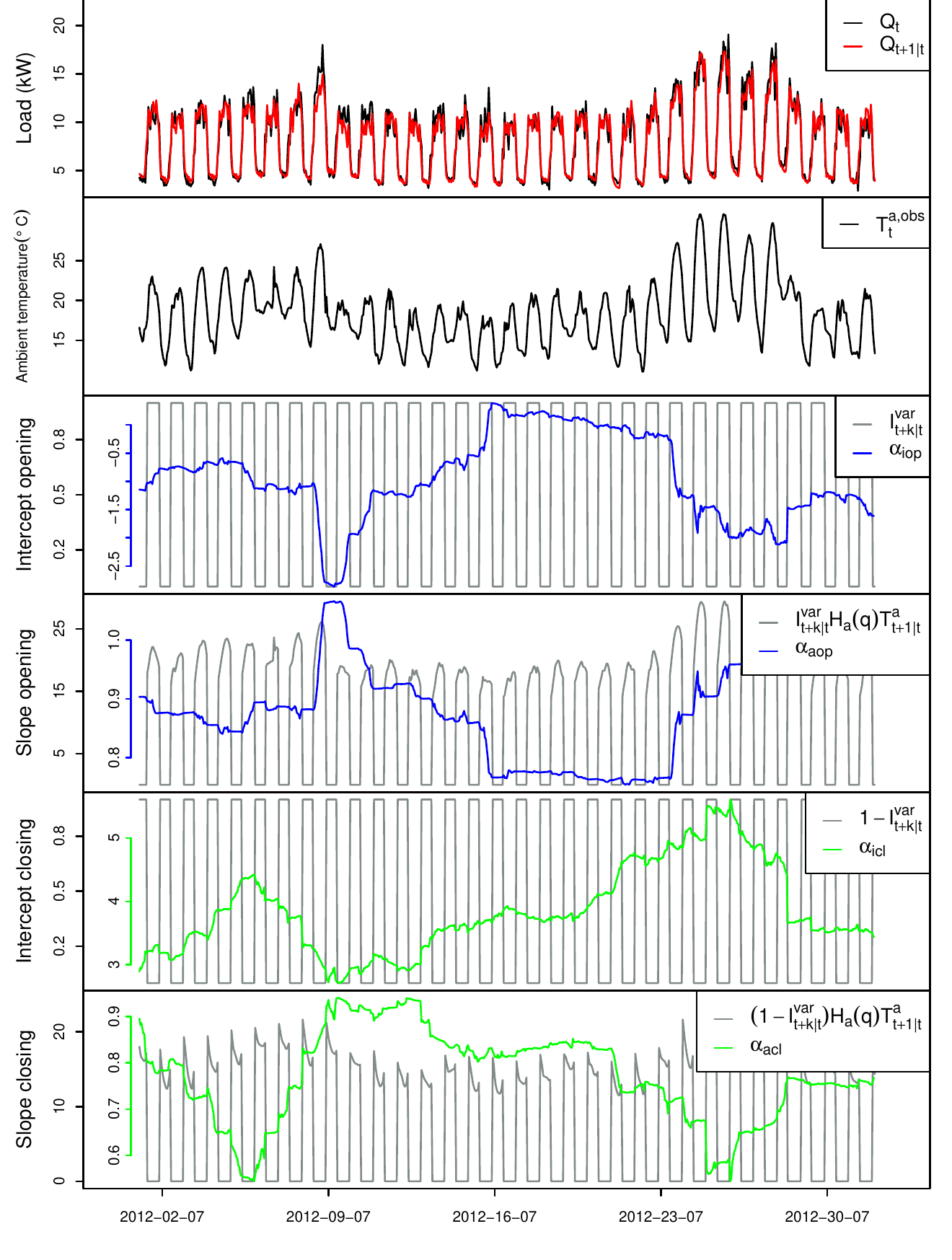} 
   \caption{The upper plot is the measured and one-step ahead forecasted load. Below this is a plot of the observed ambient temperature. The two plots in the middle are the active inputs in the opening regime: first plot is of the intercept input (grey line) and corresponding estimated coefficient (blue line), next plot is of the ambient temperature input (grey line) and corresponding estimated coefficient (blue line). The lower two plots are similar, but for the closing regime.  }
   \label{fig:coef}
\end{figure}
The behaviour of the fitted coefficients for the splined and low-pass filtered ambient temperature inputs are analysed by plotting them for July, together with the forecast and measured load and the observed temperature in Figure \ref{fig:coef}. From this plot it is seen that the slope coefficients are adapting to changes in the ambient temperature. It is seen in both in the closing and opening regime, very clearly in the opening regime during the 8'th of July. The intercepts are normally changing opposite of the slope. It is seen that the coefficient values for the slope are not far from one, which indicates that the spline functions describe the non-linear effect well. \\

\begin{figure} 
   \includegraphics[scale=1]{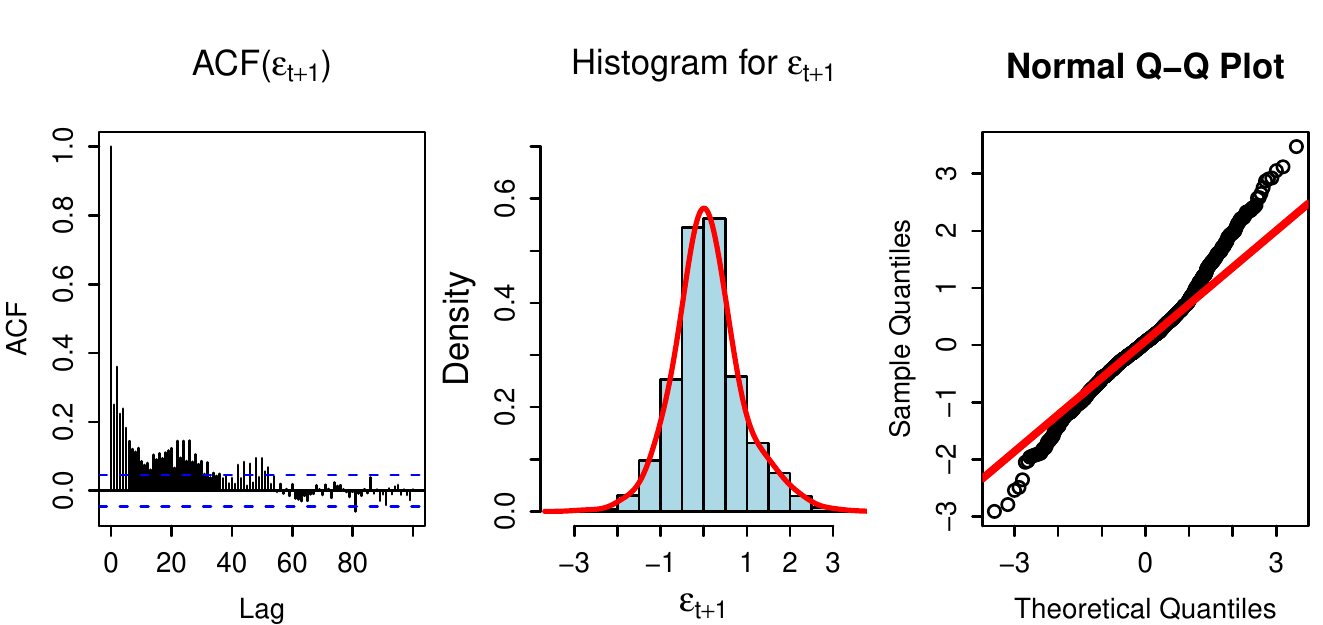} 
   \caption{The ACF, histogram and QQ-normal plot of the one-step ahead residuals from $Model_{\mathit{var.nonlin}}$.}
   \label{fig:ACF}
\end{figure}
In Figure \ref{fig:ACF} the auto-correlation function (ACF), histogram and QQ-normal plot for the one-step ahead residuals are shown. In the ACF plot a significant correlation for the shorter lags is seen. The histogram shows a non-skewed bell shaped distribution, which indicates normal distributed residuals, however the QQ-normal plot shows that the tails of the distribution are a bit heavy compared to the normal distribution. The ACF indicates that some un-described information remains and therefore the auto-regressive (AR) noise model
\begin{align}
  \epsilon_{t+1} = \alpha_{noise,1} \epsilon_{t} + \alpha_{noise,2} \epsilon_{t-1} + \alpha_{noise,24} \epsilon_{t-23} + \epsilon_{t+1}^{\mathrm{noise}}
\end{align}
is applied and fitted with RLS scheme optimizing the forgetting factor by plotting the RMSE$_1$ for steps of $0.001$ and finding $\lambda=0.997$ as the optimal value. This is the AR model with fewest lags having residual not significantly different from white noise.
\begin{figure} 
   \includegraphics[scale=1]{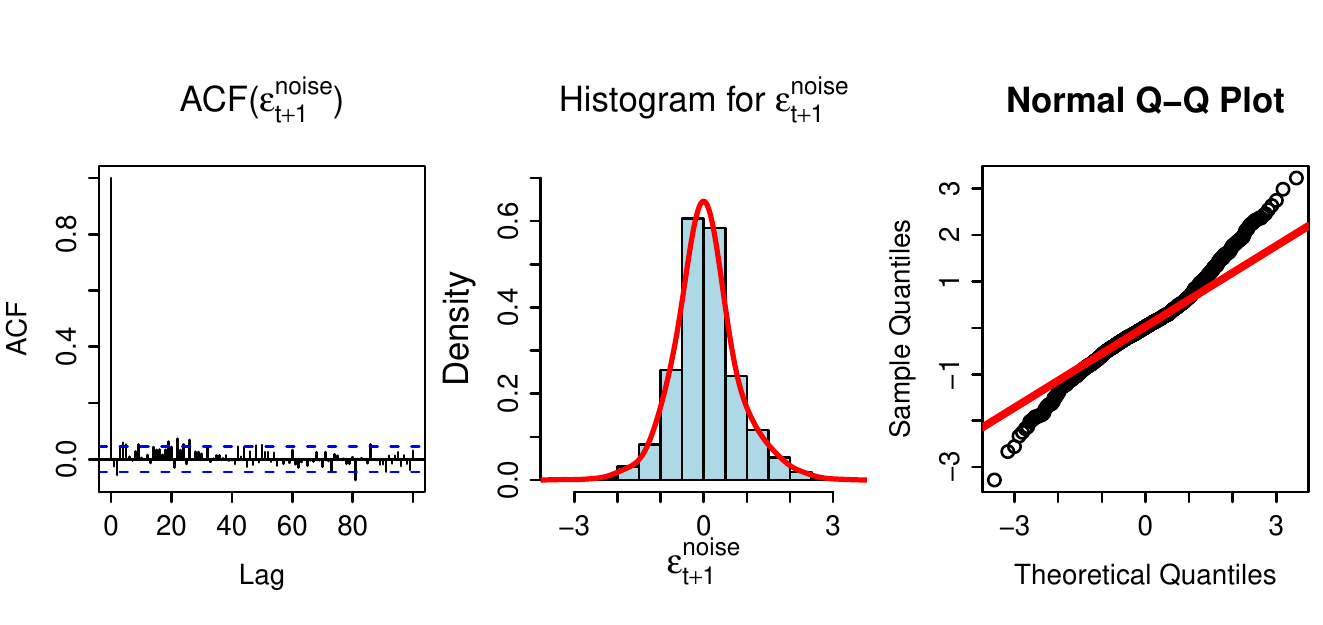} 
   \caption{The ACF, histogram and QQ-normal plot of the one-step ahead residuals from $Model_{\mathit{var.nonlin}}$ after the noise model.}
   \label{fig:ACFNoise}
\end{figure}
The ACF, histogram and QQ-normal plot for the noise model residuals are displayed in Figure \ref{fig:ACFNoise}. From the ACF it is seen that the residuals are not significantly different from white noise (assuming no non-linear dependency), however the QQ-normal plot reveals that the tails are still a bit too heavy for the residuals to normal distributed.

\section{Discussion}
\label{sec:discussion}
It is found that the applied methods and models performs very well for forecasting of the electrical load for refrigeration. The non-linearities are modelled in a two-stage approach, first using an off-line spline based regression model which then feeds into a linear time-adaptive model in a second stage. Using this approach the non-linear effects from ambient temperature are described and improves the model performance. However it is noted, that this effect must be confirmed with studies including data from a longer period, such that the first stage (the non-linear part) of the model can be fitted to historic data only.\\

A possible physical explanation to this dependency above around 17-20 $^\circ$C, could be that if the indoor air temperature is normally kept at a level around 20 $^\circ$C by a climate system, but on warm sunny days the indoor air temperature increases to above 20 $^\circ$C and then the heat transfers of the cabinets to the surroundings increases. It could also be a decrease in the coefficient of performance (COP) of the compressor rack on warm days.\\

In the report \citep{ESO} the load is found to have highly non-linear dependency with the relative humidity in the supermarket, therefore implementing the local humidity as input could be considered as input to the models. Another possible improvement could be to optimize the coefficient for the low-pass filtering separately for each regime, in order to include different dynamical relations in the regimes.\\

Further work could also be focusing on modelling the forecast uncertainties, since this will give valuable information for operation of the energy system with a high level of fluctuating renewable energy production. Furthermore, the models should be tested on other supermarkets to see how well they adapt to different systems and conditions, and to further confirm the results especially with regards to the non-linear effects.\\

\section{Conclusion}
\label{sec:conclusion}
A method for forecasting electrical load for refrigeration in a Supermarket is presented. It issues load forecast from 1 to 42 hours ahead. Load measurements and local measured ambient temperature for the period May 1$^{st}$ to August 1$^{st}$ 2012 have been used as basis for the modelling. Three models are presented, which all are formed by adaptive linear time series modelling techniques using local observations and weather forecasts as input. The models are formed by a diurnal curve and low-pass filtered ambient temperature input. In the first model the effect of ambient temperature is linear and switched between two regimes at a fixed time interval each day. The second model has a more adaptive regime switching using of a predicted diurnal curve. The third model has additionally a first stage in which a non-linear spline function is applied for modeling non-linear effects of ambient temperature. It is shown, by comparing the RMSE$_k$ for all models, that the third model performs better than the two first models due to the inclusion of the non-linear effect of ambient temperature, however it is also noted that this should be confirmed by studies in which the first stage is applied to historic values only. Finally, a thorough analysis of the residuals shows that after applying an auto-regressive noise model the one-step ahead residuals are not significantly different from white noise indicating that only little further improvement will be possible.

\section*{Acknowledgement}
Acknowledgements are given to the Danish Energy Technology Development and
Demonstration Programme (project EUDP-I ESO2) and the Danish Council for Strategic Research and the Danish Council for Technology and Innovation (project iPower), which have provided the financial support for the work. The Danish Meteorological Institute is thanked for making their numerical weather predictions available.

\bibliographystyle{model2-names}
\bibliography{litterature.bib}







\end{document}